\documentclass[amsmath,floatfix,twocolumn,superscriptaddress]{revtex4-1}
\usepackage{amssymb,amsmath}
\usepackage{graphicx,array}
\usepackage{dcolumn}
\usepackage{psfrag}
\usepackage{bm}
\usepackage{color}

\begin{document}
\title{First-principles high-throughput screening of bulk piezo-photocatalytic materials for sunlight-driven hydrogen production}

\author{Zhao Liu}
\affiliation{School of Materials Science and Engineering, Dongguan University of Technology,
Dongguan 523808, China}
\affiliation{Research Institute of Interdisciplinary Science, Dongguan University of
Technology, Dongguan 523808, China}

\author{Biao Wang}
\affiliation{School of Materials Science and Engineering, Dongguan University of Technology,                           
Dongguan 523808, China}
\affiliation{Research Institute of Interdisciplinary Science, Dongguan University of
Technology, Dongguan 523808, China}

\author{Dewei Chu}
\affiliation{School of Materials Science and Engineering, UNSW Sydney, Sydney, NSW 2052, Australia}

\author{Claudio Cazorla}
\affiliation{Departament de F\'isica, Universitat Polit\`ecnica de Catalunya, Campus Nord B4-B5, Barcelona E-08034, Spain}

\maketitle

{\bf Finding cost-effective and efficient photocatalytic materials able to catalyse the water splitting reaction 
under visible light is one of the greatest challenges in current environmental material science. Despite that many 
photocatalysts are already known in the context of green hydrogen production, strategies to systematically and 
rationally modify their optoelectronic properties to achieve desired photocatalytic performance are yet to be 
established. Piezoelectric materials react to mechanical stimuli by adjusting their band gaps and band alignments, 
thus offering a possible route to precise photocatalyst design. However, piezo-photocatalysts are relatively 
scarce and have been seldom investigated to date. Here, we present a high-throughput screening of piezo-photocatalytic 
materials performed over $\sim 1,000$ bulk piezoelectrics that relies on a simple electrostatic model and first-principles 
calculations. A total of $\sim 10$ previously overlooked binary and tertiary bulk compounds are theoretically identified 
as highly promising piezo-photocatalysts due to their appropriate optoelectronic properties and superb band alignment 
tunability driven by uniaxial strain.}
\\

About three quarters of the hydrogen consumed annually worldwide, currently around $70$ million tonnes, 
are produced by methane reforming, which generates about $830$ million tons of CO$_{2}$ emissions \cite{iea2019,idriss20}.
Aimed at curbing carbon emissions and motivated by the increasingly reduced costs of renewable electricity, 
there is growing interest in producing hydrogen from water electrolysis. However, generating the global 
hydrogen demand entirely from electricity would require an electricity consumption of $3600$ TWh, which 
is not efficient from an energy resources point of view (e.g., it surpasses the annual electricity production
of the European Union). Finding alternative means to produce clean hydrogen is therefore critical for attaining 
sustainable energy and industry futures. In this context, photocatalytic hydrogen production via solar water 
splitting emerges as one of the most promising solutions. 

\begin{figure*}[t]
\centerline{
\includegraphics[width=1.00\linewidth]{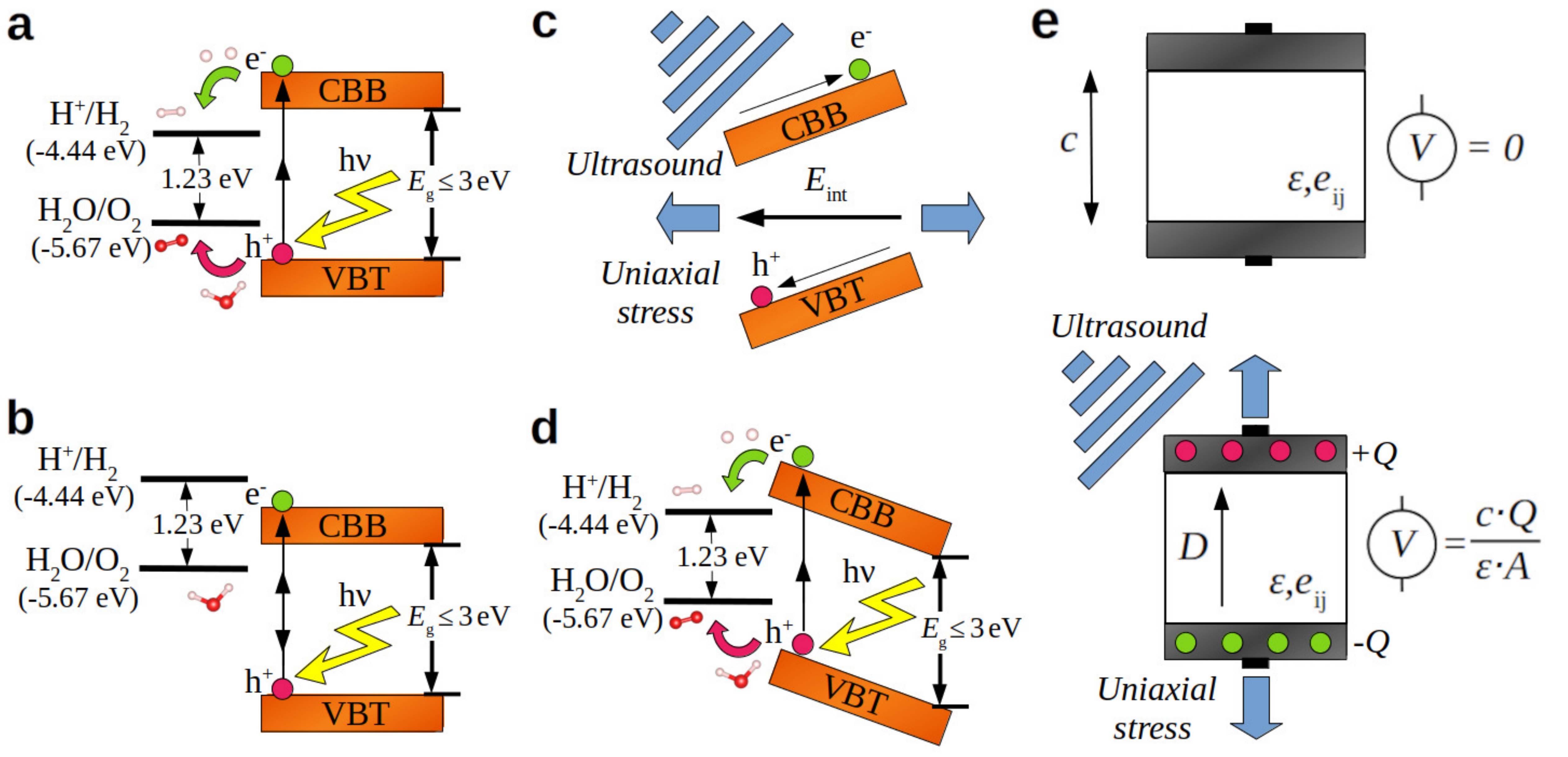}}
\caption{Fundamental aspects of photocatalytic, piezoelectric and piezo-photocatalytic materials.
         {\bf a} Opto-electronic properties of an ideal water-splitting photocatalyst; the
         conduction band bottom (CBB) and valence band top (VBT) of the photocatalyst perfectly
         straddle the OER and HER potentials as referred to the vacuum level. {\bf b} Opto-electronic
         properties of a non-ideal water-splitting photocatalyst; the CBB and VBT of the photocatalyst
         fail to straddle the OER and HER potentials. {\bf c} Application of uniaxial stress or
         ultrasound waves on a piezoelectric material induces an internal electric field and thus
         the appearance of a potential gradient. {\bf d} Improvement of OER and HER straddling in a
         non-ideal photocatalyst that is piezoelectric achieved through uniaxial stress or ultrasound
         waves. {\bf e} Simple parallel plate capacitor in which the dielectric material between
	 the plates is a piezoelectric; upon application of uniaxial stress or ultrasound waves a
	 finite voltage is created.}
\label{fig1}
\end{figure*}

Splitting water molecules into oxygen and hydrogen under visible light is possible with the assistance of 
photocatalytic materials \cite{kudo09,chen17}. Efficient photocatalytic materials should fulfill a number of 
stringent physical conditions like (Fig.\ref{fig1}a--b): (1)~the energy band gap, $E_{g}$, should be larger than 
$1.23$~eV but smaller than $\approx 3.0$~eV in order to be able to absorb the visible radiation from sunlight, 
(2)~the bottom~(top) of the conduction~(valence) band, CBB~(VBT), should be higher~(lower) than the reduction~(oxidation) 
potential H$^{+}$/H$_{2}$~(H$_{2}$O/O$_{2}$), HER~(OER), which lies at $-4.44$~($-5.67$)~eV with respect to 
the vacuum level, and (3)~the recombination rate of light-induced electron-hole charge carriers should be low.
In addition to the above requirements, potential candidates should be also abundant and easy to synthesize. 
Therefore, finding suitable photocatalytic materials remains a challenging task.

Several chemical and nanostructuring approaches have been successful at improving the photocatalytic activity 
of archetypal inorganic compounds like TiO$_{2}$ and CeO$_{2}$ under sunlight \cite{asahi01,mofarah19,bahmanrokh20,xu21}. 
However, the usual intricacy of nanostructuring methods along with the tremendous variability of the as-synthesized 
nanomaterials makes it difficult to identify general approaches for consistently enhancing the photocatalytic 
performance of crystals \cite{park16}. Consequently, progress in photocatalysis, and in particular in photocatalytic 
production of hydrogen, aided by rational design of materials remains limited \cite{takanabe17}.

Recently, some of us have shown by means of first-principles calculations that mechanical strains (biaxial and uniaxial) 
can modify the band-gap properties of transition-metal oxides in a significant and consistent manner \cite{liu20,liu21}. 
In practice, biaxial strains of $1$--$5$\% can be introduced in crystals by growing thin films on top of substrates 
that present a lattice mismatch \cite{zhang19,hu17}. Similarly, uniaxial strains of comparable sizes can be realized 
in microstructured materials via mechanical actuation \cite{dang21,phan17} and/or exposure to ultrasound waves 
\cite{blagov08,truell69}. Therefore, in view of the condition (1) above, mechanical stress in principle could be 
exploited to improve the photocatalytic activity of non-ideal materials by finely tuning their optoelectronic 
properties. However, achieving an adequate degree of predictability in the band-alignment variations of traditional 
semiconductors as induced by strain [see the condition (2) above] appears to be extremely challenging \cite{liu20}. 
Such a lack of accomplishment critically hinders rational design of photocatalysts based on strain engineering.  

Notwithstanding such limitations, a particular class of materials called piezoelectrics exhibit quantifiable and 
systematic band structure variations driven by mechanical stress. In response to a strain deformation, a finite
electric field develops in the interior of piezoelectrics that (i)~introduces an approximately linear variation 
of the energy bands throughout the crystal, and (ii)~reduces the recombination rate of migrating electron-hole pairs 
[see Fig.\ref{fig1}c and conditions (2) and (3) above]. Moreover, piezoelectrics are non-centrosymmetric materials that 
encompass all pyroelectrics and ferroelectrics, hence they are abundant. These qualities are all very desirable for 
the rational engineering of novel materials with improved optoelectronic and photocatalytic properties (Fig.\ref{fig1}d) 
\cite{wu16} and some recent research have indeed focused on the study of archetypal ``piezo-photocatalysts'' like 
ZnO and BaTiO$_{3}$ \cite{tu20,liang19,sakthivel19}. Yet, a quantitative characterization and understanding of 
piezo-photocatalytic materials is lacking and the number of piezo-photocatalysts known to date is very limited. 

In this work, we present a comprehensive first-principles computational study based on density functional theory (DFT) 
calculations that provides (i)~a simple but physically insightful description of piezo-photocatalytic materials in terms
of a piezoelectric plate capacitor (PPC) model, and (ii)~a high-throughput screening of piezo-photocatalytic materials 
performed over a large database of $\sim 1,000$ bulk piezoelectrics \cite{mp-piezo} that relies on an easy-to-compute 
DFT descriptor. (Two-dimensional materials -- e.g., g-C$_{3}$N$_{4}$ and GeS -- and solid solutions -- e.g., 
A$_{x}$A'$_{1-x}$B$_{y}$B'$_{1-y}$O$_{3}$ -- are not included in the inspected database \cite{mp-piezo} hence despite 
of their photocatalytic promise such families of compounds have been disregarded in the present study.) Based on our 
first-principles computational sieve, a number of previously overlooked binary and tertiary bulk compounds are identified 
as promising piezo-photocatalysts among which nitride (LaN and GaN), halide (PtF$_{4}$ and AgI), chalcogenide (BiTeCl, 
BiTeI, MgTe, CdS and ScCuS$_{2}$) and other inorganic (SiC) materials stand out due to their outstanding band-alignment 
tunability driven by uniaxial strain. Therefore, the present computational study has the potential to stimulate experimental 
synthesis of new piezo-photocatalytic materials able to boost the sustainable production of hydrogen based on solar water 
splitting.

\section*{Results}
\label{sec:results}
In this section, a simple electrostatic model describing the band alignment rearrangements in a piezo-photocatalyst caused 
by uniaxial strain is first introduced along with an easy-to-compute first-principles descriptor of the resulting piezo-photocatalytic 
performance. Next, we present the main results of our high-throughput screening of piezo-photocatalytic materials performed 
over a large dataset of $\sim 1,000$ bulk piezoelectrics \cite{mp-piezo} followed by a careful quality check of the employed 
first-principles data. Finally, the outcomes of our refined high-throughput screening are confronted with explicit calculations 
of the piezo-photocatalytic performance of one of the most promising identified compounds when subjected to broad uniaxial 
strain conditions. For the sake of focus, and due to obvious limitations, other relevant aspects to photocatalytic activity 
different from band gap and band alignment features (e.g., surface molecular reactions and light absorption processes) 
mostly are left for future work. 
\\

{\bf Piezo-photocatalysts bulk modelling: The piezoelectric plate capacitor (PPC).}~To approximately model the band-alignment 
behavior of piezo-photocatalytic materials under stress, we start by examining the simple case of a parallel plate capacitor 
in which the dielectric material between the plates is piezoelectric (Fig.\ref{fig1}e). Under open-circuit conditions and 
in the absence of mechanical stresses, the voltage drop among the plates of the capacitor, $V$, is null (we assume the 
spontaneous polarisation of the dielectric material to be zero). However, when a mechanical stress is applied on the 
capacitor $V$ will be different from zero as a result of the accumulation of polarisation charges of different signs in the 
two capacitor plates. For the sake of simplicity, let us assume that (i)~the mechanical stress ($\sigma$) is uniaxial 
and applied along the capacitor stacking direction, and (ii)~the dielectric piezoelectric material is isotropic (Fig.\ref{fig1}e). 
In such a case, the value of the potential drop can be evaluated as: 
\begin{equation}
	V = \frac{c \cdot Q}{\epsilon \cdot A} = \frac{c}{\epsilon} D~, 
	\label{eq1}
\end{equation}
where $c$ represents the thickness of the capacitor, $\epsilon$ the dielectric constant of the piezoelectric material and
$D$ the charge per unit area accumulated in the plates. The value of the surface charge density can be determined via the 
isothermal piezoelectric stress constant of the dielectric material defined as:
\begin{equation}
	e_{33} = \frac{\partial D}{\partial \eta}~,
        \label{eq2}
\end{equation}
where $\eta \equiv \Delta c / c$ represents the stress-induced strain. Under small deformations it can be reasonably assumed 
that $D \approx e_{33} \eta$, hence the internal electric field that appears within the piezoelectric, $E_{\rm int}$, can be 
also expressed in the linear approximation as:
\begin{equation}
	E_{\rm int} \left( \eta \right) = \frac{1}{c} V\left(\eta\right) = \frac{1}{c} \cdot \frac{\partial V}{\partial \eta} \eta \approx \frac{e_{33}}{\epsilon} \eta ~.
        \label{eq3}
\end{equation}
The following piezoelectric voltage coefficient can be defined from the formula above: 
\begin{equation}
	\alpha_{\eta} \equiv \frac{\partial V}{\partial \eta} = \frac{c}{\epsilon} e_{33}~,
\label{eq4}
\end{equation}
which describes the internal electrostatic potential variation induced by uniaxial strain. Based on the simple piezoelectric 
plate capacitor (PPC) model introduced here, it can be reasonably argued that to a first approximation the maximum band-alignment 
change induced by uniaxial strain on a piezoelectric thin film is:
\begin{equation}
	\Delta V_{i} \left(\eta \right) \equiv V_{i}\left(\eta \right)- V_{i}\left(0\right) = -c E_{\rm int}\left(\eta\right) =  -\alpha_{\eta} \eta~,
	\label{eq5}
\end{equation}
where $i$ stands for either the conduction band bottom (CBB) or valence band top (VBT) energy levels. (Here, surface dipoles 
are assumed to not vary by effect of small $\eta$'s.) 

From Eq.(\ref{eq5}), it is readily shown that piezoelectric materials presenting large (small) $\alpha_{\eta}$ coefficients 
[Eq.(\ref{eq4})] will offer great (poor) band-alignment tunability as driven by uniaxial strain. Consequently, piezoelectric 
materials possessing energy band gaps in the range of $1.23 \lesssim E_{g} \lesssim 3.0$~eV (to absorb the sunlight visible 
radiation), small dielectric constants and large piezoelectric stress coefficients (the last two conditions as for maximizing 
$\Delta V_{i}$), \emph{a priori} should be regarded as promising piezo-photocatalysts. Moreover, large $\alpha_{\eta}$ coefficients
imply large strain-induced internal electric fields [see Eqs.(\ref{eq3})--(\ref{eq4})] which are desirable for depleting 
the recombination rate of light-induced excitons (Fig.\ref{fig1}c). It is noted that large piezoelectric stress constants typically are accompanied 
by also large dielectric constants \cite{dash21,cazorla15}, thus it is naturally difficult to find materials with large band-alignment 
piezo tunability (i.e., large $\alpha_{\eta}$ values). This small set of conditions are not only physically insightful but also 
computationally convenient: the relevant quantities $E_{g}$, $\epsilon$ and $e_{33}$ can be efficiently estimated via bulk 
first-principles DFT calculations \cite{vasp,cazorla15a,cazorla17a}. As we will show in the next section, this circumstance 
can be exploited to conduct high-throughput computational searches of piezo-photocatalysts within large databases of piezoelectric 
materials that are publicly available \cite{mp-piezo}. 

It is important to note that the simple PPC model introduced here in principle should not be regarded as quantitavely accurate 
due to the involved simplifications (e.g., strain-induced $E_{g}$ variations and surface dipole and surface relaxation effects 
have been neglected, and the assumed isotropic piezoelectric behaviour should not always be valid). Thus, any candidate 
piezo-photocatalyst identified on basis to the $\alpha_{\eta}$ descriptor should necessarily be put to test. In the next sections, 
we will comment on the extent of the possible limitations of the PPC model introduced in this section.  
\\

\begin{figure*}[t]
\centerline{
\includegraphics[width=1.00\linewidth]{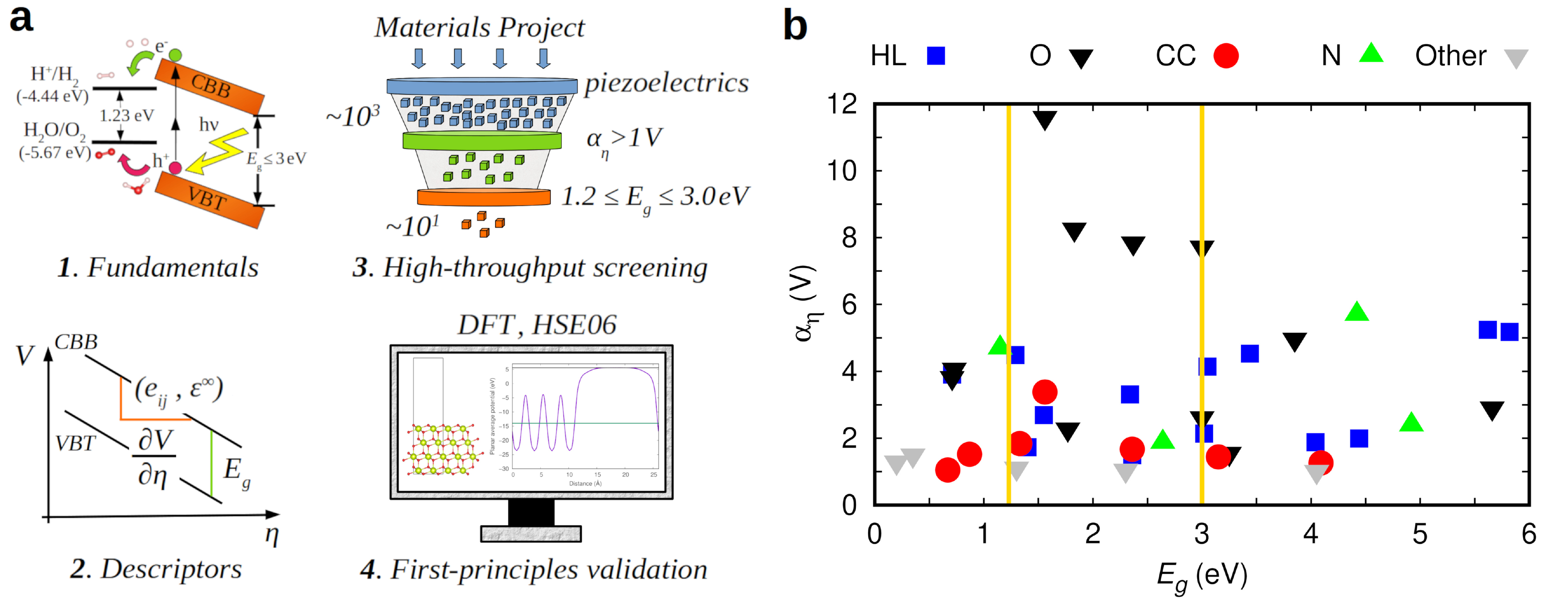}}
\caption{High-throughput screening of piezo-photocatalytic materials for water splitting under visible light: 
	strategy and results. {\bf a} The adopted high-throughput screening strategy consists of a sequence 
	of four steps. First, to understand the fundamentals of piezo-photocatalysts; second, to define a 
	suitable set of easy-to-compute descriptors for piezo-photocatalytic materials; third, based on 
	threshold descriptor values, to identify potential piezo-photocatalytic materials from the ``Materials 
	Project'' database \cite{mp-piezo}; and fourth, to validate the results of the high-throughput screening 
	by explicitly computing with first-principles methods the optoelectronic and band alignment properties 
	of some of the identified piezo-photocatalysts. {\bf b} High-throughput screening results. A total of $21$ 
	compounds were identified as potential piezo-photocatalytic materials for water splitting under visible 
	light (Table~I). The descriptor threshold values of $1.23 \lesssim E_{g} \lesssim 3.0$~eV are represented 
	with gold vertical lines in the figure and only materials displaying $\alpha_{\eta} > 1$~V are considered. 
	``HL'' stands for materials mostly containing halide atoms (F, Cl, Br, I), ``O'' oxygen atoms, ``CC'' 
	chalcogenide atoms (S, Se, Te), ``N'' nitrogen atoms, and ``Other'' refers to inorganic compounds not 
	classified in the previous categories.}
\label{fig2}
\end{figure*}

{\bf High-throughput piezo-photocatalysts screening.}~We performed a high-throughput computational search of piezo-photocatalysts
over the Materials Project (MP) database containing about $1,000$ different bulk piezoelectric compounds \cite{mp-piezo}
(Methods). Figure~\ref{fig2} shows the steps followed in our high-throughput materials screening (Fig.\ref{fig2}a) 
and the main results obtained from it (Fig.\ref{fig2}b). Based on the PPC model introduced in the previous section 
and the DFT data available in the MP database, we applied two consecutive descriptor sieves on the reported bulk piezoelectrics. 
Compounds exhibiting a $\alpha_{\eta}$ coefficient larger than $1$~V [Eq.(\ref{eq4})] were first selected (which amounted to $42$, 
see Supplementary Table~I; thus this condition encompasses the top $\approx 5$\% of all the investigated materials in terms
of band alignment tunability) and from them those fulfilling the band-gap condition $1.23 \lesssim E_{g} \lesssim 3.0$~eV were 
finally retrieved. A total of $21$ compounds out of the initial $\sim 1,000$ piezoelectrics were tentatively identified as 
promising piezo-photocatalysts (Fig.\ref{fig2}b), all of which are listed in Table~I along with their most relevant properties. 

Among the most promising piezo-photocatalysts we found oxide (e.g., KNbO$_{3}$, BaTiO$_{3}$, NaNbO$_{3}$ and 
KIO$_{3}$), nitride (e.g., LaN and Ge$_{2}$N$_{2}$O) and halide (e.g., PtF4, BaI$_{2}$, BrF$_{3}$ and SeBr) materials, 
most of which have not been previously investigated in the context of green hydrogen production. In terms of band-alignment 
tunability, non-centrosymmetric oxide perovskites (i.e., KNbO$_{3}$, BaTiO$_{3}$ and NaNbO$_{3}$) emerge as the clear 
winners since they exibit huge $\alpha_{\eta}$ values larger than or close to $10$~V. Several chalcogenides (e.g., BiTeCl 
and MgTe) and generic inorganic compounds (GaP and SiC) were also identified as potential photocatalysts, some of which 
appear to be especially promising in terms of reduced production costs (i.e., SiC). 

It is worth noting that our first-principles based high-throughput search identified already well known piezo-photocatalysts 
like BaTiO$_{3}$ \cite{tu20,liang19,sakthivel19} and KNbO$_{3}$ \cite{yu19,zhang21}. Other piezoelectric materials 
frequently employed in optoelectronic applications like ZnO and InN \cite{InN} were not retrieved by our computational 
analysis because the corresponding $E_{g}$ values appearing in the MP database were noticeably smaller than $1.23$~eV. 
For example, for hexagonal ZnO (space group $P6_{3}mc$ and MP identity number ${\rm mp-2133}$) we estimated a large 
$\alpha_{\eta}$ coefficient value of $4.07$~V, which turns out to be competitive with those of the top piezo-photocatalyst 
candidates listed in Table~I; however, the ZnO band gap reported in the MP database amounts to $0.73$~eV, which in principle 
would fail to properly straddle the OER and HER potentials \cite{liu20}. 

Nonetheless, it is worth recalling here the inherent limitations of standard DFT approaches in estimating accurate band 
gaps in semiconductors due to the ubiquitous electronic self-interaction errors and other fundamental problems \cite{perdew86}. 
In particular, it is well known that common exchange-correlation functionals like LDA and GGA, which are computationally 
very affordable, tend to significantly underestimate $E_{g}$ \cite{hse06}, and unfortunately (although also understandably) 
most of the results reported in the MP database are obtained with such standard DFT functionals. Thus, it is very likely 
that some of the candidate piezo-photocatalysts identified by our high-throughput search actually do not fulfill the band 
gap condition $E_{g} \lesssim 3.0$~eV. Similarly, it is also quite likely that some potentially good piezo-photocatalysts 
have escaped our computational sieve due to the systematic DFT underestimation of their band gap (e.g., hexagonal ZnO \cite{liu20}, 
see next section). Furthermore, the reliability of the simple PPC model introduced in this work, upon which the present 
computational search has been built, needs to be assessed. Consequently, we undertook a careful evaluation of the MP data 
and PPC model employed in our computational search by performing supplementary first-principles calculations. 
\\ 

\begin{table*}
\centering
\begin{tabular}{c c c c c c c c c}
\hline
\hline
$ $ & $ $ & $ $ & $ $ & $ $ & $ $ & $ $ & $ $ & $ $ \\
	${\rm Formula} $ & \quad  ${\rm mp-ID}$ \quad & \quad ${\rm Space~Group}$ \quad & \quad $c$ \quad & \quad $\epsilon$ \quad & \quad $|e_{33}|$ & \quad $E_{g}$ \quad & \quad $\alpha_{\eta}$ & \quad ${\rm Type}$ \\
	$ $ \quad & \quad $ $ \quad & \quad $ $ \quad & \quad $ $(\AA) \quad & \quad $(\epsilon_{0})$ \quad & \quad $({\rm C} \cdot {\rm m}^{-2})$ \quad & \quad ${\rm (eV)}$ \quad & \quad ${\rm (V)}$ & $ $\\
$ $ & $ $ & $ $ & $ $ & $ $ & $ $ & $ $ & $ $ & $ $\\
\hline
$ $ & $ $ & $ $ & $ $ & $ $ & $ $ & $ $ & $ $ & $ $\\
	${\rm KNbO_{3}}$    \qquad & ${\rm 4342}$   & $P4mm~{\rm (99)}$              & $4.09$ & $12.98$ & $3.26$ & $1.56$ & $11.6$& ${\rm O} $ \\
	${\rm BaTiO_{3}}$   \qquad & ${\rm 5986}$   & $P4mm~{\rm (99)}$              & $4.07$ & $19.24$ & $3.45$ & $1.83$ & $8.26$& ${\rm O} $ \\
	${\rm NaNbO_{3}}$   \qquad & ${\rm 4681}$   & $Pmc2_{1}~{\rm (26)}$          & $6.39$ & $29.60$ & $3.22$ & $2.37$ & $7.85$& ${\rm O} $ \\
	${\rm KIO_{3}}$     \qquad & ${\rm 552729}$ & $R3m~{\rm (160)}$              & $4.53$ & $ 9.03$ & $1.36$ & $3.00$ & $7.72$& ${\rm O} $ \\
	${\rm LaN}$         \qquad & ${\rm 567290}$ & $P6_{3}mc~{\rm (186)}$         & $4.74$ & $20.48$ & $1.79$ & $1.15$ & $4.69$& ${\rm N} $ \\
	${\rm PtF_{4}}$     \qquad & ${\rm 8943}$   & $Fdd2~{\rm (43)}$              & $6.02$ & $ 5.92$ & $0.39$ & $1.29$ & $4.49$& ${\rm HL}$ \\
	${\rm BaI_{2}}$     \qquad & ${\rm 568536}$ & $P\overline{6}2m~{\rm (189)}$  & $7.94$ & $14.52$ & $0.67$ & $3.05$ & $4.14$& ${\rm HL}$ \\
	${\rm BiTeCl}$      \qquad & ${\rm 28944}$  & $P6_{3}mc~{\rm (186)}$         & $7.58$ & $ 4.57$ & $0.18$ & $1.56$ & $3.38$& ${\rm CC}$ \\
	${\rm BrF_{3}}$     \qquad & ${\rm 23297}$  & $Cmc2_{1}~{\rm (36)}$          & $5.32$ & $19.30$ & $1.06$ & $2.34$ & $3.31$& ${\rm HL}$ \\
	${\rm SeBr}$        \qquad & ${\rm 570589}$ & $Aea2~{\rm (41)}$              & $8.02$ & $ 4.72$ & $0.14$ & $1.55$ & $2.69$& ${\rm HL}$ \\
	${\rm KNO_{3}}$     \qquad & ${\rm 6920}$   & $R3m~{\rm (160)}$              & $4.44$ & $ 4.22$ & $0.22$ & $3.00$ & $2.62$& ${\rm O} $ \\
	${\rm Na_{2}O_{2}}$ \qquad & ${\rm 2340}$   & $P\overline{6}2m~{\rm (189)}$  & $5.69$ & $ 5.36$ & $0.19$ & $1.77$ & $2.28$& ${\rm O} $ \\
	${\rm TlF} $        \qquad & ${\rm 558134}$ & $Aem2~ {\rm (39)}$             & $4.67$ & $48.33$ & $1.96$ & $3.02$ & $2.14$& ${\rm HL}$ \\
	${\rm Ge_{2}N_{2}O}$\qquad & ${\rm 4187}$   & $Cmc2_{1}~{\rm (36)}$          & $5.42$ & $ 8.18$ & $0.25$ & $2.64$ & $1.87$& ${\rm N}$ \\
	${\rm BiTeI}$       \qquad & ${\rm 22965}$  & $P3m1~{\rm (156)}$             & $5.41$ & $ 6.67$ & $0.20$ & $1.33$ & $1.84$& ${\rm CC}$ \\
	${\rm GaN}$         \qquad & ${\rm 804}$    & $P6_{3}mc~{\rm (186)}$         & $3.89$ & $11.65$ & $0.46$ & $1.74$ & $1.75$& ${\rm N} $ \\
	${\rm AgI}$         \qquad & ${\rm 22894}$  & $P6_{3}mc~{\rm (186)}$         & $5.68$ & $ 7.43$ & $0.20$ & $1.40$ & $1.73$& ${\rm HL}$ \\
	${\rm MgTe}$        \qquad & ${\rm 1039}$   & $P6_{3}mc~{\rm (186)}$         & $5.58$ & $ 8.31$ & $0.22$ & $2.36$ & $1.67$& ${\rm CC}$ \\
	${\rm PI_{3}}$      \qquad & ${\rm 27529}$  & $P6_{3}~{\rm (173)}$           & $7.79$ & $ 2.96$ & $0.05$ & $2.36$ & $1.49$& ${\rm HL}$ \\
	${\rm GaP}$         \qquad & ${\rm 8882}$   & $P6_{3}mc~{\rm (186)}$         & $4.72$ & $13.02$ & $0.27$ & $1.30$ & $1.11$& ${\rm Other}$ \\
	${\rm SiC}$         \qquad & ${\rm 7140}$   & $P6_{3}mc~{\rm (186)}$         & $3.75$ & $11.32$ & $0.28$ & $2.30$ & $1.05$& ${\rm Other}$ \\
$ $ & $ $ & $ $ & $ $ & $ $ & $ $ & $ $ & $ $ & $ $ \\
\hline
\hline
\end{tabular}
 \caption{Materials classified as potentially good piezo-photocatalysts for water splitting under visible light according
	 to our first-principles high-throughput screening. Compounds are ranked according to their $\alpha_{\eta}$ value
	 estimated with the DFT data retrieved from the Materials Project (MP) database. ``mp-ID'' stands for the compound 
	 identification number in the MP database \cite{mp-piezo}, ``HL'' for halide materials, ``O'' for oxides, ``CC'' for 
	 chalcogenides, ``N'' for nitrides, and ``Other'' for inorganic compounds not classified in the previous categories.} 
\end{table*}

\begin{figure*}[t]
\centerline{
\includegraphics[width=1.00\linewidth]{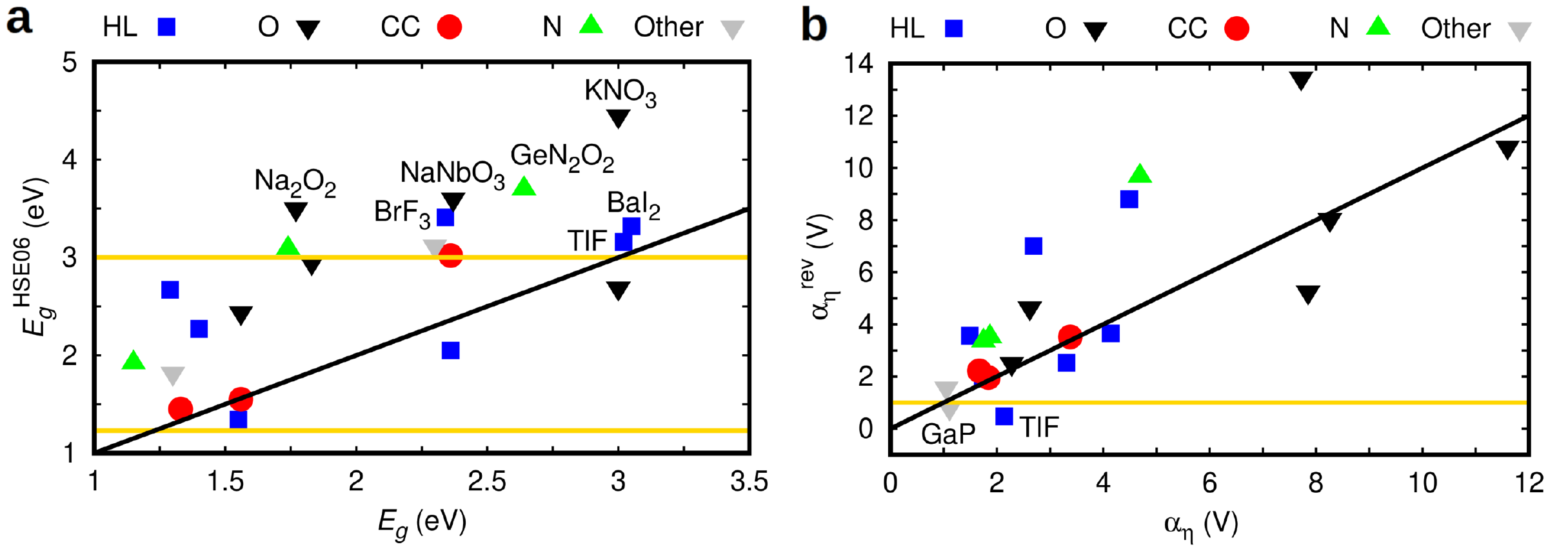}}
\caption{Refinement of the first-principles data retrieved from the MP database. {\bf a} MP band gap, $E_{g}$, versus 
	HSE06 band gap, $E_{g}^{\rm HSE06}$ \cite{hse06}. {\bf b} Piezoelectric voltage coefficient calculated from the 
	MP database, $\alpha_{\eta}$, and from fully converged first-principles calculations based on the PBEsol \cite{pbesol} 
	and HSE06 functionals, $\alpha_{\eta}^{\rm rev}$. As a result of our DFT data refinement, $8$ out of the initially 
	identified $21$ potentially good piezo-photocatalysts for water splitting under visible light (Table~I) were discarded 
	(i.e., those named in the figures) since they did not fulfill either the condition of $1.23 \lesssim E_{g}^{\rm HSE06} 
	\lesssim 3.0$~eV and/or $\alpha_{\eta}^{\rm rev} > 1$~V (gold lines). The solid black lines with a slope equal to one are 
	guides to the eye for better appreciating the $E_{g}$--$E_{g}^{\rm HSE06}$ and $\alpha_{\eta}$--$\alpha_{\eta}^{\rm rev}$ 
	discrepancies.}   
\label{fig3}
\end{figure*}

\begin{table}
\centering
\begin{tabular}{c c c c c}
\hline
\hline
$ $ & $ $ & $ $ & $ $ & $ $ \\
${\rm Formula} $ & \quad ${\rm Space~Group}$ \quad & \quad $E_{g}^{\rm HSE06}$ \quad & \quad $\alpha_{\eta}^{\rm rev}$ & \quad ${\rm Drawbacks} $ \\
$ $  &  $ $ & ${\rm (eV)}$ & ${\rm (V)}$ & $ $ \\
$ $ & $ $ & $ $ & $ $ & $ $ \\
\hline
$ $ & $ $ & $ $ & $ $ & $ $ \\
${\rm KIO_{3}}$     & $R3m~{\rm (160)}$        & $2.69$ & $13.46$ & ${\rm Water~soluble} $ \\
${\rm KNbO_{3}}$    & $P4mm~{\rm (99)}$        & $2.44$ & $10.81$ & ${\rm -} $\\
${\rm LaN}$         & $P6_{3}mc~{\rm (186)}$   & $1.92$ & $9.67$  & ${\rm -} $\\
${\rm PtF_{4}}$     & $Fdd2~{\rm (43)}$        & $2.67$ & $8.80$  & ${\rm High~cost} $ \\
${\rm BaTiO_{3}}$   & $P4mm~{\rm (99)}$        & $2.95$ & $8.03$  & ${\rm -} $\\
${\rm CdS}$         & $P6_{3}mc~{\rm (186)}$   & $2.14$ & $7.69$  & ${\rm -} $\\
${\rm SeBr}$        & $Aea2~{\rm (41)}$        & $1.34$ & $7.00$  & ${\rm Water~soluble}$\\
${\rm ScCuS_{2}}$   & $P3m1~{\rm (156)}$       & $1.42$ & $4.10$  & ${\rm -} $\\
${\rm ZnO}$         & $P6_{3}mc~{\rm (186)}$   & $2.73$ & $4.07$  & ${\rm -} $\\
${\rm PI_{3}}$      & $P6_{3}~{\rm (173)}$     & $2.05$ & $3.56$  & ${\rm Water~reacting} $\\
${\rm BiTeCl}$      & $P6_{3}mc~{\rm (186)}$   & $1.55$ & $3.52$  & ${\rm High~cost} $\\
${\rm GaN}$         & $P6_{3}mc~{\rm (186)}$   & $3.09$ & $3.35$  & ${\rm -} $\\
${\rm MgTe}$        & $P6_{3}mc~{\rm (186)}$   & $3.02$ & $2.22$  & ${\rm High~cost} $\\
${\rm AgI}$         & $P6_{3}mc~{\rm (186)}$   & $2.27$ & $1.99$  & ${\rm -} $\\
${\rm BiTeI}$       & $P3m1~{\rm (156)}$       & $1.45$ & $1.96$  & ${\rm High~cost} $\\
${\rm 2H-SiC}$      & $P6_{3}mc~{\rm (186)}$   & $3.12$ & $1.57$  & ${\rm -} $\\
$ $ & $ $ & $ $ & $ $ & $ $ \\
\hline
\hline
\end{tabular}
 \caption{Revised list of potential piezo-photocatalysts for water splitting under visible light based 
	 on the refinement of the DFT data employed in our initial first-principles high-throughput 
	 screening. Compounds are ranked according to their revised piezoelectric voltage coefficient, 
	 $\alpha_{\eta}^{\rm rev}$. Potential practical drawbacks are indicated for each material.} 
\end{table}

{\bf First-principles data refinement.}~We re-evaluated the structural, dielectric, piezoelectric and band gap properties of
the $21$ compounds listed in Table~I by using stringent convergence parameters and fairly accurate DFT functionals (Methods). 
Specifically, we re-optimized the relevant atomic geometries with a large energy cut-off of $800$~eV and a {\bf k}--point grid 
of spacing $2 \pi \times 0.01$~\AA$^{-1}$ for sampling of the first-Brillouin zone, using the PBEsol exchange-correlation 
functional \cite{pbesol} (this DFT functional has been consistently ranked among the best performers in terms of lattice parameters 
prediction for semiconductors \cite{park21}). The dielectric, piezoelectric and band gap properties of the PBEsol-relaxed structures 
were subsequently estimated with the range-separated hybrid HSE06 potential \cite{hse06}. By proceeding in this manner, an homogeneous 
and consistently high level of numerical accuracy was guaranteed for all the investigated materials.

Figure~\ref{fig3}a shows a comparison of the band gaps found in the MP database, $E_{g}$, and those estimated by following the 
strategy explained above, $E_{g}^{\rm HSE06}$, for all the materials reported in Table~I. As it was already expected, for most 
compounds $E_{g}^{\rm HSE06}$ turns out to be significantly larger than $E_{g}$ (discrepancies amount to $60$--$110$\% in the 
worse cases; see deviations from the solid black line in Fig.\ref{fig3}a). Upon correction of the usual band-gap underestimation 
obtained with local and semi-local DFT methods, $7$ out of the $21$ initially selected potential piezo-photocatalysts were 
discarded because their $E_{g}^{\rm HSE06}$'s were noticeably larger than $3$~eV. The rejected materials were: NaNbO$_{3}$, 
Na$_{2}$O$_{2}$, BrF$_{3}$, GeN$_{2}$O$_{2}$, KNO$_{3}$, BaI$_{2}$ and TlF (Fig.\ref{fig3}a). It is noted that few compounds 
exhibiting energy band gaps marginally larger than $3$~eV were not discarded (i.e., GaN, MgTe and SiC) since our estimations 
may still contain small numerical imprecisions of the order of $0.1$~eV \cite{perdew86}. 

Likewise, Fig.\ref{fig3}b shows a comparison of the piezoelectric voltage coefficients calculated with the DFT data found in 
the MP database, $\alpha_{\eta}$, and the corresponding revised values estimated as explained above, $\alpha_{\eta}^{\rm rev}$,
for all the compounds reported in Table~I. In this occasion, the consensus between the two sets of equivalent data is somewhat 
improved as compared to the band-gap case, although from a quantitative point of view their agreement is still far from satisfactory 
(in few instances the discrepancies are larger than $100$\%; see deviations from the solid black line in Fig.\ref{fig3}b). 
Interestingly, we appreciate a quite systematic underestimation of $\alpha_{\eta}^{\rm rev}$ obtained with local and semi-local 
DFT approaches, which for our present materials screening purposes does not represent a critical shortcoming (i.e., since we are 
interested in finding compounds with piezoelectric potential coefficients larger than a specific threshold value). Nevertheless, 
in the particular case of TlF and GaP we found that $\alpha_{\eta}^{\rm rev} < 1$~V hence they were discarded as potential 
piezo-photocatalysts (it is worth noting that upon refinement of the MP data only TlF failed to fulfill the two conditions guiding 
our computational searches).

Given the typical $E_{g}$ underestimation obtained with standard DFT approaches, in our data refinement we also considered those 
materials that, based on the available MP data, presented energy band gaps below $1.23$~eV and fulfilled the condition 
$\alpha_{\eta} > 1$~V (Supplementary Table~I). Consequently, a total of $9$ additional compounds were picked up, among which only 
$3$ passed our two-conditions sieve upon refinement of their data. Those three added potential piezo-photocatalysts are the 
oxide ZnO and the chalcogenides CdS and ScCuS$_{2}$. 

Therefore, as a result of our supplementary first-principles calculations, a total of $8$ out of the $21$ initially identified 
candidate materials were rejected while $3$ out of the $21$ initially rejected compounds were recovered, adding up to a total of 
$16$ potentially good piezo-photocatalysts. This definitive set of promising compounds is listed in Table~II along with their 
re-evaluated $E_{g}^{\rm HSE06}$ and $\alpha_{\eta}^{\rm rev}$ values (additional materials properties can be found in the 
Supplementary Table~II). There are appreciable quantitative differences among the two rankings of materials shown in Tables~I 
and II that will be commented in the Discussion section. Next, we turn our attention to first-principles validation of the proposed 
PPC model and the computationally sieved piezo-photocatalytic materials. 
\\

\begin{figure*}[t]
\centerline{
\includegraphics[width=1.00\linewidth]{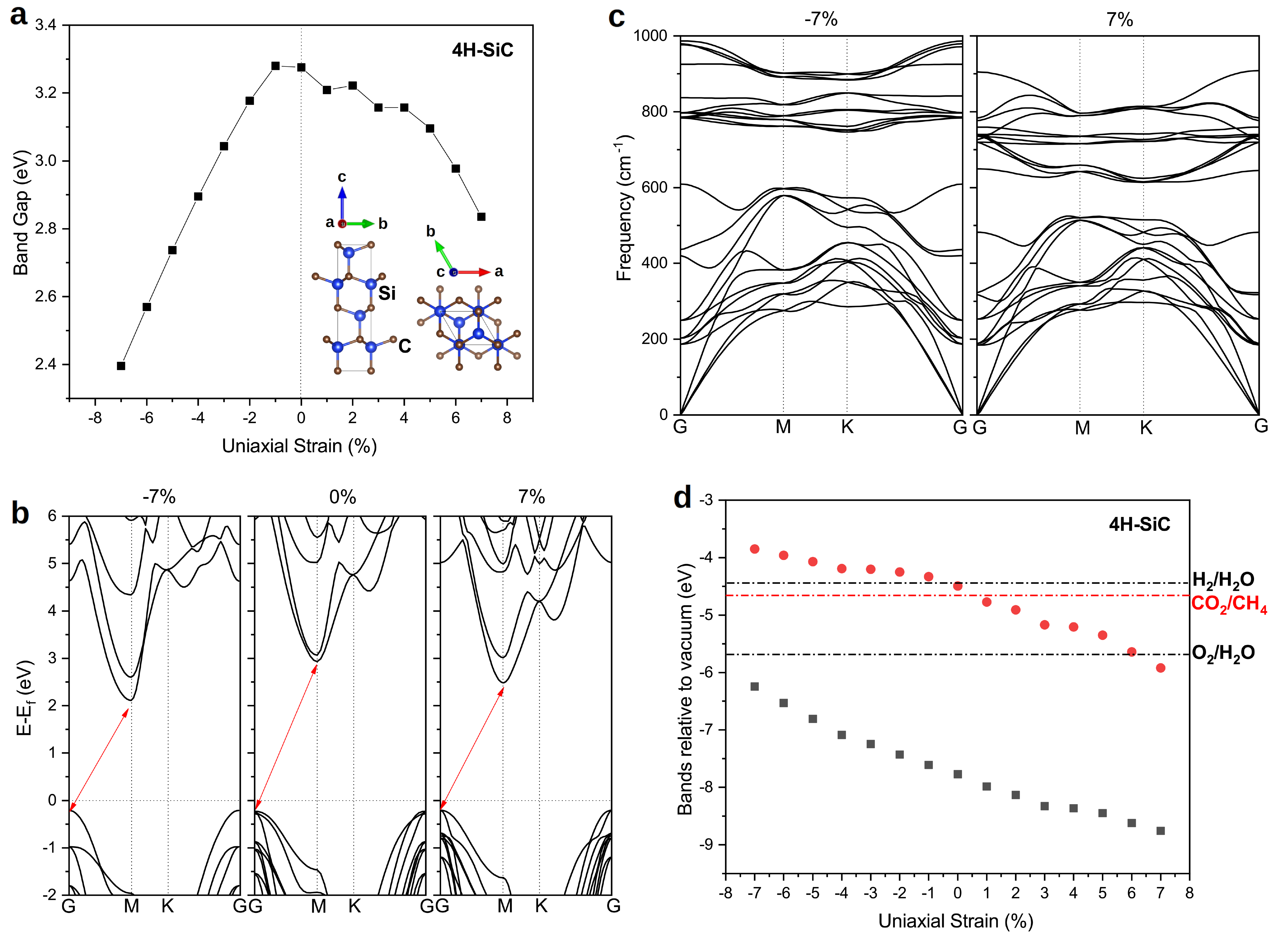}}
\caption{First-principles analysis of the optoelectronic, vibrational, and band alignment properties of piezoelectric 4H-SiC.
        {\bf a} Band-gap variation induced by uniaxial strain. {\bf b} Dependence of the electronic band structure 
	on uniaxial strain and reciprocal space point. {\bf c} Phonon frequencies estimated as a function of 
	uniaxial strain and reciprocal space point. {\bf d} Effect of uniaxial strain on the band alignments of the 
	crystal as referred to the vacuum level. The VBT and CBB energy levels are represented with solid grey 
	squares and red circles, respectively.}
\label{fig4}
\end{figure*}

\begin{figure*}[t]
\centerline{
\includegraphics[width=1.00\linewidth]{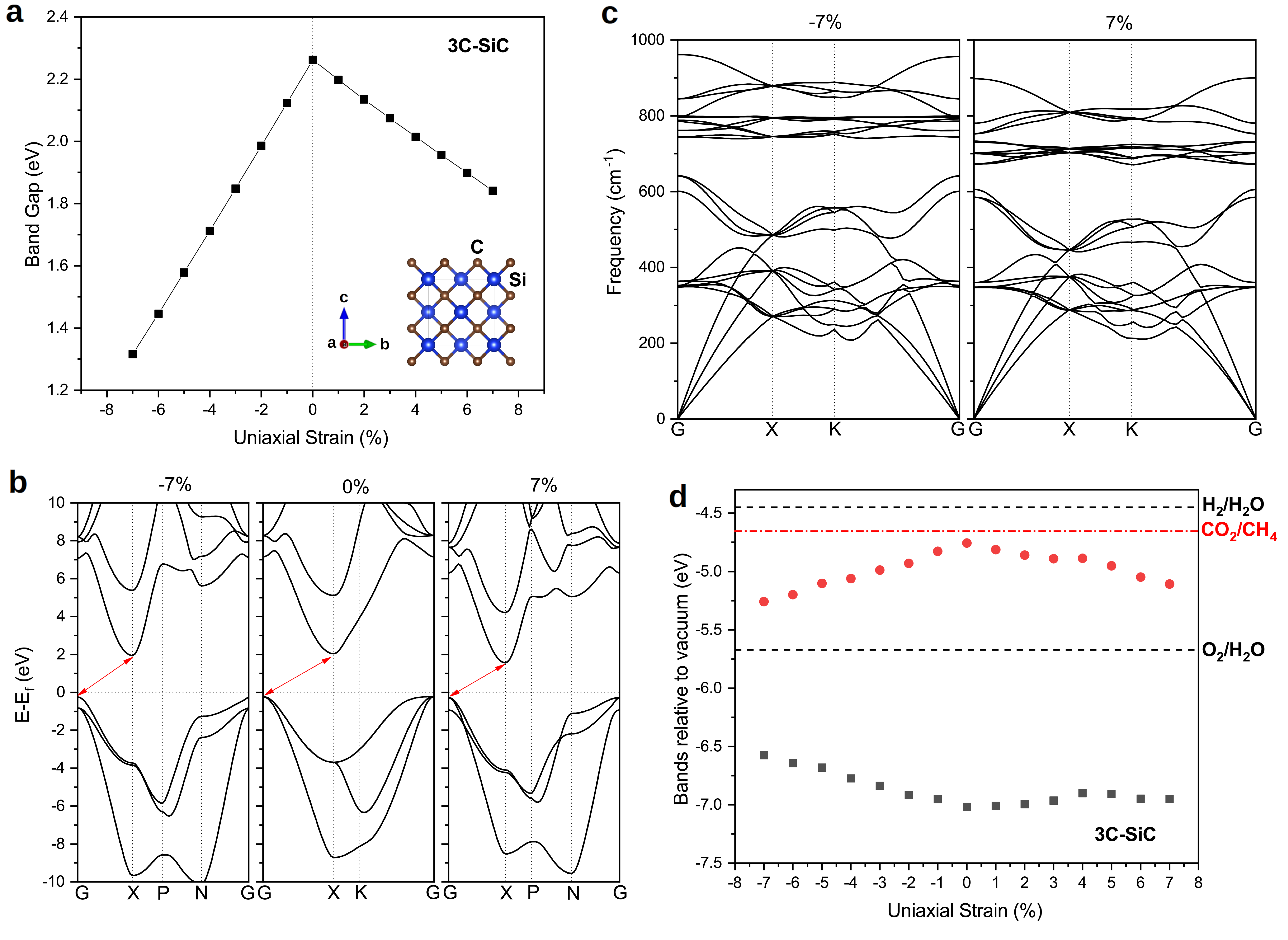}}
\caption{First-principles analysis of the optoelectronic, vibrational, and band alignment properties of non-piezoelectric 
	3C-SiC. {\bf a} Band-gap variation induced by uniaxial strain. {\bf b} Dependence of the electronic band structure 
        on uniaxial strain and reciprocal space point. {\bf c} Phonon frequencies estimated as a function of 
        uniaxial strain and reciprocal space point. {\bf d} Effect of uniaxial strain on the band alignments of the 
        crystal as referred to the vacuum level. The VBT and CBB energy levels are represented with solid grey 
        squares and red circles, respectively.}
\label{fig5}
\end{figure*}

{\bf First-principles validation of the PPC model and identified piezo-photocatalysts.}~For this crucial and computationally 
intensive part of our study, namely, explicit first-principles assessment of piezo-photocatalytic performances 
(i.e., not relying on materials descriptors) and testing of the PPC model, we selected SiC for several reasons. 
First, it is a very well-known and chemically simple material made of abundant and non-toxic elements that 
can be synthesised in a commercially large scale, hence from an applied point of view it is a very promising 
compound. Second, SiC exhibits several polymorphs some of which are piezoelectric (e.g., 2H and 4H) some 
of which are not (e.g., 3C), thus comparing their photocatalyst performances under uniaxial strain may be 
particularly meaningful. Third, despite of its high mechanical strength, several works in the literature have 
reported successful stabilisation of tensile uniaxial strains of $\approx 7$~\% in nanostructured SiC \cite{phan17},
which is the maximum $|\eta|$ considered in the present work. Fourth, previous experimental works have already shown 
that unstrained SiC is a promising photocatalytic material for driving the water splitting reaction \cite{liu12,yasuda12,wang15,wang15b}. 
And fifth, SiC appears in the last position of the lists enclosed in Tables~I and II hence positive evaluation 
of this compound in principle may also suggest suitable piezo-photocatalytic performances for the rest of materials 
ranked above it. 

Figure~\ref{fig4}a shows the impact of uniaxial strain on the band gap of SiC considering the piezoelectric
polymorph 4H (hexagonal, space group $P6_{3}mc$). Strain-induced band lifting effects are fully taken into 
consideration in our first-principles DFT results. In the absence of any stress, the $E_{g}^{\rm HSE06}$ of 
4H-SiC is indirect (Fig.\ref{fig4}b) and approximately amounts to $3.3$~eV. (It is noted that the SiC band gap 
reported in Table~II corresponds to the 2H polymorph, which is slightly smaller; for our validation analysis 
we have selected 4H-SiC because this is one of the most well-known and easy to synthesise silicon carbide polymorphs.) 
Under either tensile or compressive uniaxial strains, $E_{g}^{\rm HSE06}$ is significantly reduced and its 
indirect nature is conserved (Fig.\ref{fig4}b). The maximum $\eta$-induced band-gap reduction is achieved in 
the compressive side, where appropriate photocatalytic values below $3$~eV are obtained for uniaxial strains 
larger than $\approx 3$\% (in absolute value). Meanwhile, in the tensile side the strain-induced $E_{g}^{\rm HSE06}$ 
variation is less regular and suitable photocatalytic band-gap values are only attained at uniaxial distorsions 
larger than $\approx 5$\%. We have checked that the vibrationally stability of the 4H-SiC polymorph is preserved 
upon application of the largest strains considered in this study (i.e., no imaginary lattice phonon bands appear
in the two limit cases $\eta = \pm 7$\%, Fig.\ref{fig4}c). 

Figure~\ref{fig4}d shows the variation of the 4H-SiC band alignments as induced by uniaxial strain (Methods).
At zero strain, the VBT and CBB levels fail to correctly straddle the OER and HER potentials due to 
the existence of large negative offsets (e.g., the VBT level is approximately positioned $2$~eV below 
the OER energy). When uniaxial strain is applied, however, the VBT and CBB levels are significantly 
displaced towards higher energies in the compressive side and towards lower energies in the tensile side. 
As a result, upon compressive uniaxial strains of $|\eta| > 4$\% the band alignments of 4H-SiC are optimally 
positioned for straddling the OER and HER levels. It is also appreciated that the induced band-alignment 
changes can be regarded as roughly linear in $\eta$ (at least locally). Therefore, based on our first-principles 
calculations, it is concluded that although unstrained 4H-SiC does not seem to be an optimal water-splitting 
photocatalyst its green hydrogen production performance can be substantially improved by means of compressive 
uniaxial strain (i.e., both the band gap and band alignments can be appropriately tuned for such an end). 
Analogous band-alignment results are also obtained for the piezoelectric 2H-SiC polymorph (hexagonal, space 
group $P6_{3}mc$), although in this latter case straddling of the OER and HER levels is already attained in 
the absence of structural distortions (Supplementary Fig.1).

The outcomes obtained for 4H-SiC are encouraging and qualitatively consistent with the simple PPC model 
introduced above, that is, in agreement with the potential energy variation predicted by Eq.(\ref{eq5}). 
At the quantitative level, however, the accordance between the computed $\alpha_{\eta}$ coefficients 
and the explicitly estimated first-principles $dV/d\eta$ variations are fair but not exact. For instance, by 
taking numerical derivatives on the VBT data represented in Fig.\ref{fig4}d we obtain a $\eta$-induced 
potential variation of $29~(6)$, $18~(6)$ and $16~(6)$~V for uniaxial strains of $-7$, $0$ and $+7$\%, 
respectively (numerical uncertainties are expressed within parentheses). Meanwhile, the physically equivalent 
$\alpha_{\eta}$ values calculated with Eq.(\ref{eq4}) for the same lattice strains are $22$, $1$ and $23$~V, 
respectively (i.e., the structural, dielectric and piezoelectric parameters defining the $\alpha_{\eta}$ 
coefficient have been recalculated at each uniaxial strain). Clearly, the numerical agreement between the 
two sets of $\alpha_{\eta}$ data is not satisfactory although in the two limit $|\eta|$ cases, for which the 
$e_{33}$ coefficient largely increases (and so does $\alpha_{\eta}$), the accordance is significantly improved. 
Several straightforward reasons explaining the lack of quantitative accuracy of the PPC model are: (i)~the 
neglection of $\eta$-induced band gap variations (i.e., the VBT and CBB levels are treated identically whereas 
in practice they may react slightly different to strain, as it is shown in Fig.\ref{fig4}d); (ii)~the assumption 
of a perfectly isotropic piezoelectric medium (i.e., in practice, most piezoelectric tensors 
exhibit non-zero off-diagonal components hence the disregarded secondary lattice distortions may also accumulate 
charge along the principally strained direction and so influence the potential variation in the material);
and (iii) the disregarding of surface dipole and surface relaxation effects. 

It is worth commenting on the sign of the $\Delta V_{i}$ shift ($i =$ VBT, CBB) as induced by $\eta$. In standard 
piezoelectrics, piezoelectric and piezoelectric voltage (Eq.\ref{eq4}) coefficients are positively defined 
($e_{33}, \alpha_{\eta} > 0$); thus, according to the PPC model, under tensile uniaxial strain ($\eta > 0$) 
the electrostatic potential in the interior of the material should rise as referred to the zero vacuum level (Eq.\ref{eq5} 
and Fig.\ref{fig1}c). Conversely, under compressive unaxial strain ($\eta < 0$) the electrostatic potential in the 
interior of the piezoelectric should come closer to the zero vacuum level (Eq.\ref{eq5}). These $\Delta V_{i}$ trends 
predicted by the PPC model in fact are clearly reproduced by the DFT results enclosed in Fig.\ref{fig4}d. Furthermore, 
we performed supplementary band alignment calculations for an anomalous piezoelectric material exhibiting negative 
piezoelectric and piezoelectric voltage coefficients ($e_{33}, \alpha_{\eta} < 0$). In this latter case, the 
expected $\eta$-induced $\Delta V_{i}$ shifts are opposite to those just explained, namely, tensile (compressive) 
uniaxial strain reduces (increases) the electrostatic potetial in the interior of the material as referred to the 
zero vacuum level. Consistently, our DFT calculations reproduce the potential trends anticipated by Eq.(\ref{eq5}) 
also in this case (Supplementary Fig.2). 

Figure~\ref{fig5} shows equivalent results to those just explained for 4H-SiC but obtained for the non-piezoelectric 
polymorph 3C-SiC (cubic, space group $F\overline{4}3m$). The band gaps are also indirect in this case but noticeably 
smaller than estimated for 4H-SiC (Fig.\ref{fig5}a,b). (Strain-induced band lifting effects are fully taken into 
consideration in our first-principles DFT results.) For instance, in the absence of strain $E_{g}^{\rm HSE06}$ amounts 
to $2.3$~eV and at $\eta = -7$\% to $1.3$~eV. The $\eta$-induced band-gap variations in 3C-SiC resemble those calculated 
for 4H-SiC but they are more regular and somewhat larger. Likewise, the vibrational stability of the system is not 
affected by the lattice strain (Fig.\ref{fig5}c). Nonetheless, the band alignments of non-piezoelectric 3C-SiH under 
strain behave radically different from those of piezoelectric 4H-SiC (Fig.\ref{fig5}d). In particular, the induced VBT 
and CBB shifts are quite small (i.e., at most $\approx 0.5$~eV) and not proportional to the applied strain (i.e., 
essentially they seem to follow the $\eta$-driven $E_{g}^{\rm HSE06}$ variations found in both compressive and 
tensile sides). Consequently, the water-splitting photocatalytic properties of non-piezoelectric 3C-SiC cannot be 
effectively tuned via uniaxial stress and the system remains largely unsuitable for green hydrogen production (i.e., 
appropriate straddling of the OER and HER energy levels is never achieved through $\eta$).

\begin{figure*}[t]
\centerline{
\includegraphics[width=1.00\linewidth]{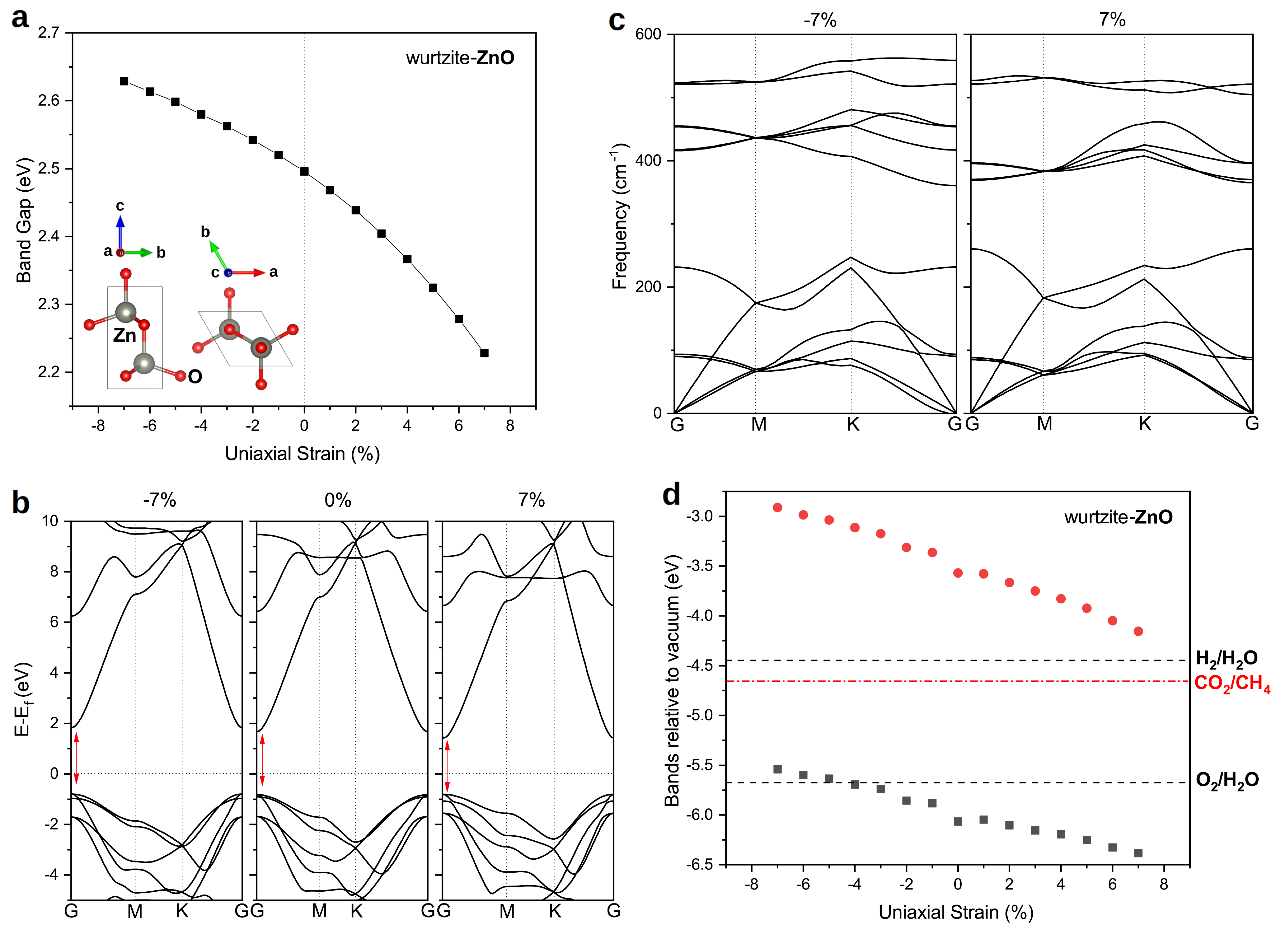}}
\caption{First-principles analysis of the optoelectronic, vibrational, and band alignment properties of wurtzite
        ZnO. {\bf a} Band-gap variation induced by uniaxial strain. {\bf b} Dependence of the electronic band
        structure on uniaxial strain and reciprocal space point. {\bf c} Phonon frequencies estimated as a function
        of uniaxial strain and reciprocal space point. {\bf d} Effect of uniaxial strain on the band alignments
        of the crystal as referred to the vacuum level. The VBT and CBB energy levels are represented with solid grey
        squares and red circles, respectively.}
\label{fig6}
\end{figure*}

For the sake of completeness and motivated by the fact that previous experimental and computational 
studies have shown that ZnO is a promising water-splitting photocatalytic material \cite{liu20,tu20,liang19,sakthivel19,hamid17,zhang18,cao22}, 
we explicitly estimated the $\eta$-induced band-gap and band-alignment variations for the wurtzite 
phase of this material. Figure~\ref{fig6} shows the results of our first-principles calculations carried 
out for piezoelectric ZnO (hexagonal symmetry, space group $P6_{3}mc$). Strain-induced band lifting effects 
are fully taken into consideration in our first-principles DFT results. Under (compressive) tensile uniaxial 
strain the band gap of wurtzite ZnO remains direct and its size steadily (increases) decreases 
(Fig.\ref{fig6}a-b). The estimated $E_{g}$ variation is less pronounced than found in SiC (Figs.\ref{fig4}-\ref{fig5}), 
although more regular. For example, the band-gap reduction (increase) attained at the highest tensile 
(compressive) strain considered here is of $\approx 11$\% ($\approx 6$\%). At the same time, the vibrational 
stability of wurtzite ZnO is pertinently conserved under uniaxial strain (Fig.\ref{fig6}c). The influence 
of uniaxial strain in the band alignments of ZnO (Fig.\ref{fig6}d) is qualitatively very similar to that 
found in 4H-SiC (Fig.\ref{fig4}d). In particular, both the VBT and CBB levels are shifted towards more 
(less) negative potential values under tensile (compressive) strains and the induced potential variations 
are quasi linear. The $dV/d\eta$ values estimated for wurtzite ZnO, however, are roughly two times smaller 
than those found in 4H-SiC. Interestingly, according to our calculations the VBT and CBB positions in 
unstrained ZnO are already quite optimal for driving the watter splitting reaction hence very small 
structural distortions should be enough to optimally adjust its piezo-photocatalyst performance. 

Several general conclusions can be extracted from the results presented in this section. First, the band alignments 
of piezoelectric materials in fact can be tailored by uniaxial strain in a systematic and efficient manner. Second, 
the $\eta$-induced band-alignment changes predicted by the simple PPC model are physically well-motivated and thus 
can be used for the screening of potential piezo-photocatalytic materials. (Actually, whenever the band alignments 
of a piezoelectric compound are known the simple PPC model can be used to roughly estimate the amount of tensile or 
compressive uniaxial strain that is necessary to shift them towards optimal levels.) And third, the list of potential 
water-splitting piezo-photocatalytic materials enclosed in Table~II should be regarded as trustworthy, hence it may 
be key in guiding strain-assisted experiments in the field of green hydrogen production.

\section*{Discussion}
\label{sec:discussion}
The two rankings of piezo-photocatalytic materials enclosed in Tables~I and II, respectively deduced from the 
DFT data found in the MP database and our refined DFT calculations, display appreciable quantitative differences. 
For instance, in Table~I the value of the reported piezoelectric potential coefficients are typically smaller 
than in Table~II and the top positions in the first list are predominantly occupied by oxide perovskites (whereas 
KIO$_{3}$ in the first position of Table~II is an inorganic salt). Furthermore, upon refinement of the structural, 
piezoelectric and dielectric DFT data several materials initially classified as suitable water-splitting piezo-photocatalysts 
in Table~I were finally discarded and not included in Table~II (e.g., KNO$_{3}$, GaP and TlF), whereas few materials 
that initially were discarded finally were included in Table~II (i.e., ZnO, CdS and ScCuS$_{2}$). On the other hand, 
at the qualitative level both materials rankings are generally equivalent since oxide, nitride and halide compounds 
are consistently presented as the most promising. Therefore, for piezo-photocatalyst screening purposes the DFT data 
contained in the MP database should be regarded as of sufficient quality. For the rest of this section, however, 
we will concentrate our analysis on the materials reported in Table~II.

The first of all ranked piezo-photocatalysts, KIO$_{3}$, appears to be auspicious both in terms of band  
alignments tunability (i.e., largest $\alpha_{\eta}^{\rm rev}$ value) and economical costs (i.e., all the 
integrating elements are abundant); however, this material is soluble in water and therefore is of little 
practical relevance in the context of water splitting photocatalysis. Similar water solubility and water 
reactivity drawbacks are posed by the compounds SeBr and PI$_{3}$ (Table~II), which should neither be 
contemplated in the context of green hydrogen production applications. Notwithstanding such limitations, 
these water unsuited piezo-photocatalytic materials could turn out to be useful for catalyzing other green 
chemistry reactions not requiring of aqueous media (e.g., CO$_{2}$ reduction \cite{jiang20}) hence we 
cautiously keep them in our list of selected candidates. 

In the second position of Table~II, it appears the perovskite oxide KNbO$_{3}$ for which recently its great 
piezo-photocatalytic potential has been experimentally demonstrated \cite{yu19,zhang21}. In fact, according 
to our DFT calculations KNbO$_{3}$ possesses superior photocatalytic band structure features than BaTiO$_{3}$
(which appears in the fifth position), namely, a slightly smaller band gap and a larger $\alpha_{\eta}^{\rm rev}$ 
coefficient. Likewise, ZnO, another known photocatalytic and piezo-photocatalytic compound, shows up in the 
ninth position of our definitive list, which comes to assure the realiability of our computational screening 
method.  

\begin{figure*}[t]
\centerline{
\includegraphics[width=1.00\linewidth]{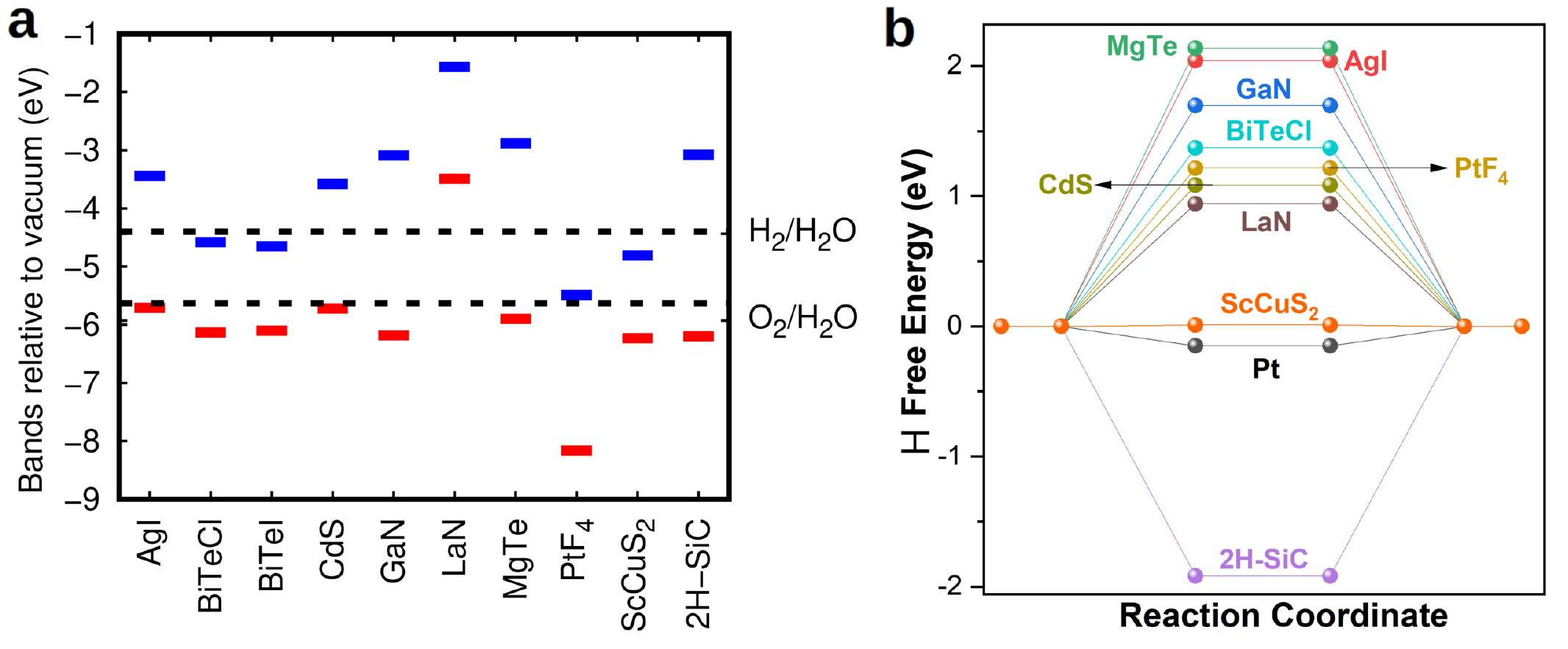}}
\caption{First-principles estimation of the {\bf a} band alignments and {\bf b} hydrogen adsorption free energy 
	of the most promising piezo-photocatalytic materials coming out from our computational screening in the 
	absence of uniaxial strain.}
\label{fig7}
\end{figure*}

In the third and fourth positions of Table~II, we find LaN and PtF$_{4}$, two materials that, to the best of 
our knowledge, have not been investigated thus far as potential water splitting photocatalysts in spite of their 
appropriate band gap and large band alignments tunability. PtF$_{4}$, however, contains precious elements hence 
its main drawback may be the high costs associated with its manufacturing. Similar expense limitations may also 
affect the three chalcogenide compounds BiTeCl, MgTe and BiTeI appearing in the eleventh, thirteenth and fifteenth 
positions of Table~II, respectively, since they contain the rare-earth element Te. Concerning these latter systems, 
it may be interesting to explore whether similar piezo-photocatalytic performances can be attained by means of 
chemical substitutions based on the more abundant chalcogenide species Se and S. 

The two chalcogenides CdS and ScCuS$_{2}$ show up in the sixth and eighth positions of Table~II, respectively.
CdS is an already known semiconductor material in the context of photocatalytic hydrogen production since it 
presents suitable band edge positions and a band gap that is responsive to visible light \cite{kudo09}. Actually, 
CdS becomes an excellent photocatalyst for H$_{2}$ evolution under visible light irradiation when aided by hole 
scavenger co-catalysts that help to prevent possible photocorrosion issues \cite{nasir20}. Thus, in view of its
high ranking position in Table~II, right after BaTiO$_{3}$, and already appropriate photocatalytic features in 
the absence of mechanical stimulation, CdS emerges as a highly promising piezo-photocatalytic compound. ScCuS$_{2}$, 
on the other hand, is a relatively unknown material that only recently has been investigated in the context of 
possible thermoelectric and photovoltaic applications by means of first-principles simulation methods \cite{rugut21,scanlon10}.
Nevertheless, according to our DFT calculations, this material is highly auspicious for photocatalysis due to 
its great band alignment tunability under strain and favorable hydrogen evolution activity (see below).

Furthermore, GaN, a well-known direct band-gap semiconductor extensively used in light-emitting applications, 
appears in the twelfth position of our ranking. The photocatalytic activity of this material in water-splitting 
hydrogen production has been already assessed and seems to be quite promising \cite{kida06,shimosako21}. Since
GaN is a well-known piezoelectric and technologically relevant material, it is likely that its piezo-photocatalytic 
performance will be eventually assessed; however, to the best of our knowledge, this possibility has not been 
yet addressed in the scientific literature. On this regard, our high-throughput computational screening suggests 
that it could be beneficial to consider also the alike nitride LaN since this compound ranks in the third position 
of our candidate piezo-photocatalysts list and in terms of production costs seems to be similar to GaN.  

By overlooking the water soluble and reacting (i.e., KIO$_{3}$, SeBr and PI$_{3}$), high cost (i.e., PtF$_{4}$, 
BiTeCl, MgTe and BiTeI) and already known (i.e., KNbO$_{3}$, BaTiO$_{3}$ and ZnO) piezo-photocatalytic compounds 
reported in Table~II, we are left with six promising candidates: the nitrides LaN and GaN, the chalcogenides CdS 
and ScCuS$_{2}$, the halide AgI and the semiconductor SiC. Most of these materials (i.e., all except ScCuS$_{2}$) 
are already known from their use in numerous applications ranging from industry and electronics to energy conversion, 
and they can be efficiently produced in different morphologies and large quantities (e.g., AgI and SiC \cite{she17,aznar17}). 
Therefore, even if the piezoelectric coefficient of some of them are noticeably smaller that those of, for instance, 
KNbO$_{3}$ and PtF$_{4}$, it is certainly worth assessing their actual water-splitting piezo-photocatalytic 
qualities in experiments. 

The overall photocatalytic performance of a piezoelectric material in fact does not solely depend on its band gap
and band alignment tunabilities, which are the two main aspects that have been considered throughout the present work. Other 
features like the initial band positions in the unstrained state, the recombination rate of photogenerated carriers 
and surface overpotentials for the hydrogen and oxygen evolution reactions, for instance, may be also determinant. 
Theoretical evaluation of some of these properties, however, involve cumbersome calculations hence we leave them for 
future work. Nevertheless, for the sake of completeness and to motivate experimental efforts, we evaluated the zero-strain 
band alignments and hydrogen evolution energetics of the candidate piezo-photocatalysts listed in Table~II, excepting 
those materials that pose water-related issues (i.e., KIO$_{3}$, SeBr and PI$_{3}$) and were already known (i.e., 
KNbO$_{3}$, BaTiO$_{3}$ and ZnO). 

Figure~\ref{fig7}a shows the band edge positions of the screened candidate piezo-photocatalysts estimated at zero strain. 
Except for PtF$_{4}$ and LaN, the obtained results appear to be all very encouraging. For instance, compounds AgI, CdS, 
GaN, MgTe and 2H-SiC satisfactorily straddle the OER and HER potentials whereas BiTeCl, BiTeI and ScCuS$_{2}$ practically 
succeed in doing so. In all these cases, therefore, employing strain strategies to adjust their band edge positions to optimal 
values appears to be fully justified. 

Figure~\ref{fig7}b shows the calculated free energy of an hydrogen atom adsorbed on the surface of the candidate 
piezo-photocatalysts, $\Delta G_{\rm H}$ (Methods), a common descriptor employed for the evaluation of HER activity 
\cite{gao18,uosaki16}. An ideal HER catalyst, like Pt, should interact very weakly with hydrogen and thus render 
$\Delta G_{\rm H}$ values close to zero. According to our DFT calculations, ScCuS$_{2}$ is a superb HER material since  
the corresponding $\Delta G_{\rm H}$ is smaller than that of Pt and practically null. In order of decreasing HER performance, 
LaN, CdS, PtF$_{4}$, BiTeCl and GaN follow after ScCuS$_{2}$. MgTe, AgI and 2H-SiC, on the other hand, seem to interact too strongly 
with hydrogen hence, without further modifications, HER activity may be inhibited on the surface of these materials. Interestingly, we found that in some 
cases $\Delta G_{\rm H}$ can be significantly changed by uniaxial strain (Supplementary Fig.3). Specifically, in 
PtF$_{4}$ and ScCuS$_{2}$ the sign of the H adsorption free energy can be reversed by means of modest $\eta$'s, thus 
suggesting the possibility of making $\Delta G_{\rm H}$ arbitrarily small by selecting the right strain. In 
2H-SiC, we also found that $\Delta G_{\rm H}$ can be reduced by roughly $25$\% of its size via application of a $5$\% 
tensile strain.

\section*{Conclusions}
\label{sec:conclusions}
We have presented a first-principles high-throughput screening of potential bulk piezo-photocatalytic materials 
able to accelerate the production of green hydrogen from water under sunlight when exposed to ultrasound waves 
and/or mechanical actuation. Our computational sieve relies on the DFT information found in the Materials Project 
database, which comprises about $1,000$ bulk piezoelectrics, and an easy-to-compute bulk material descriptor 
deduced from a simple ``piezoelectric plate capacitor'' electrostatic model. In short, ideal piezo-photocatalysts 
should simultaneously exhibit large piezoelectric stress coefficients and small dielectric constants, two qualities 
that typically oppose each other, besides reasonable band gaps and VBT and CBB energy levels in the unstrained state. 
Already known good piezo-photocatalysts were retrieved by our computational searches (i.e., KNbO$_{3}$, BaTiO$_{3}$
and ZnO) while some other compounds previously overlooked in the context of photocatalysis were also identified as 
very promising. Specifically, it was found that, in terms of strain-driven band-alignment tunability, the piezo-photocatalytic 
water splitting ability of the nitrides LaN and GaN, the halides PtF$_{4}$ and AgI, the chalcogenides BiTeCl, BiTeI,
MgTe, CdS and ScCuS$_{2}$ and the semiconductor SiC may be comparable to those of BaTiO$_{3}$ and ZnO. Other potential 
piezo-photocatalysts like bulk KIO$_{3}$ and PI$_{3}$ were also recognized in spite of their inappropriateness to work 
in aqueous environments (hence in practice are not suitable for water-splitting photocatalytic applications). The 
introduced piezo-photocatalyst screening approach is general and can be also applied to other important families of 
materials not considered in this study (e.g., two-dimensional compounds and solid solutions). Therefore, the present 
computational work advances knowledge in the field of state-of-the-art photocatalytic materials and is expected to 
motivate original and prolific experimental investigations in the context of sustainable hydrogen production driven 
by sunlight.

\section*{Methods}
\label{sec:methods}
{\bf First-principles calculations.}~First-principles calculations based on density functional theory (DFT) \cite{vasp,cazorla15a,cazorla17a} 
were performed to simulate and analyze the influence of uniaxial strain, $\eta$, on the optoelectronic, vibrational
and band alignment properties of piezo-photocatalytic materials. The PBEsol functional \cite{pbesol} was used as 
it is implemented in the VASP software package \cite{vasp}. We employed the ``projector augmented wave'' method to 
represent the ionic cores \cite{paw} by considering the most relevant electrons of each atomic species as valence. 
Wave functions were represented in a plane-wave basis truncated at $800$~eV. For integrations within the Brillouin 
zone (BZ) of all materials we employed Monkhorst-Pack {\bf k}--point grids \cite{kpoint} of spacing $2 \pi \times 0.01$~\AA$^{-1}$. 
Uniaxially strained bulk geometry relaxations were performed with a conjugate-gradient algorithm that allowed for 
volume variations while imposing the structural constraints defining uniaxial strain \cite{liu21}. In our simulations, 
uniaxial strain is defined as $\eta = \left(c - c_{0}\right)/c_{0}$, where $c_{0}$ represents the length of the 
strained lattice vector in the absence of any stress. Positive $\eta$ values are considered tensile uniaxial strains 
and $\eta < 0$ compressive. Periodic boundary conditions were applied along the three directions defined by the 
lattice vectors hence possible surface effects were completely neglected in the bulk simulations (not so in the slab 
simulations, see below). The relaxations were halted when the forces acting on the atoms were all below $0.005$~eV$\cdot$\AA$^{-1}$. 
By using these technical parameters we obtained zero-temperature energies that were converged to within $0.5$~meV per 
formula unit. Uniaxial strain conditions were simulated at $\Delta \eta = 1$\% intervals. In order to estimate accurate 
dielectric, piezoelectric and band gap properties, we employed the range-separated hybrid HSE06 exchange-correlation 
functional \cite{hse06} to perform single-point calculations on the equilibrium geometries determined at the PBEsol 
level \cite{shenoy19}. 
 
To estimate phonon frequencies we employed the ``small-displacement'' approach \cite{phon}, in which the force-constant 
matrix of the crystal is calculated in real space by considering the proportionality between the atomic displacements 
and forces when the former are sufficiently small (in the present study this condition was satisfied for atomic 
displacements of $0.02$~\AA). Large supercells containing hundreds of atoms were employed to guarantee that the 
elements of the force-constant matrix presented practically negligible values at the largest atomic separations.  
The computation of the nonlocal parts of the pseudopotential contributions were performed in reciprocal space in 
order to maximise the numerical accuracy of the computed forces. Once a force-constant matrix was determined, we 
Fourier transformed it to obtain the phonon frequencies for any arbitrary ${\bf k}$-point in the first BZ. This latter 
step was performed with the PHONOPY code \cite{phonopy}, in which the translational invariance of the system was 
exploited to ensure that the three acoustic branches were exactly zero at the $\Gamma$ point. Central differences 
for the atomic forces, that is, both positive and negative atomic displacements, were considered. 

To calculate accurate band alignments we followed the work done by Moses and co-workers on binary semiconductors 
\cite{moses11}. Briefly, both bulk and slab calculations were performed from which the alignment of the electrostatic 
potential within the semiconductor material could be obtained relative to the vacuum level. From the slab calculations, 
the difference between the average electrostatic potential within the semiconductor material and in vacuum was obtained. 
From the bulk calculations, the band structure shifts relative to the average electrostatic potential were determined. 
These calculations were performed at each $\eta$ point and involved the estimation of macroscopic and planar average 
potentials. The planar potential was computed by averaging potential values within a well defined plane (for instance, 
perpendicular to the surface of the slab), and the macroscopic potential was obtained by taking averages of the planar 
potential over distances of one unit cell along the chosen direction \cite{resta88,cazorla12}. The slab systems were 
thick enough to ensure that the electron density in the center of the slab was practically equal to that in the bulk 
material. We found that $\approx 2.0$~nm thick semiconductor slabs (e.g., around $8$ SiC layers) accompanied by similarly 
large portions of vacuum provided sufficiently well converged results for the electrostatic potential and surface energy 
(Supplementary Fig.4). Band alignments were systematically estimated at the geometrical center of the slabs. A standard 
hydrogen passivation scheme \cite{shiraishi90} was employed in the geometry relaxations of the wurtzite ZnO slabs in 
order to appropriately determine the relevant macroscopic and planar average potential levels (Supplementary Fig.5).  
\\

{\bf High-throughput screening.}~The high-throughput screening of potential piezo-photocatalyst was performed using the 
Matminer package \cite{matminer}. A pre-existing dataset containing about $1,000$ piezoelectric materials was loaded 
from the Materials Project database \cite{mp-piezo}. Based on our PPC model, the parameters needed for the high-throughput 
screening were the thickness ($c$), dielectric constant ($\epsilon$) and piezoelectric constant ($e_{33}$) of each considered 
compound (Results section). The piezoelectric constants of such materials were retrieved from the piezoelectric tensors 
provided in the same dataset, and their thicknesses were approximated by the mean average of the three corresponding 
lattice parameters. The value of the dielectric constants and energy band gaps ($E_{g}$) were acquired from the Materials 
Project (MP) database via the corresponding materials mp-ID's; compounds for which the dielectric properties were not 
reported in the MP database were dropped from our high-throughput screening. Therefore, it is possible that few piezoelectric 
materials with potentially high piezo-photocatalytic tunability were overlooked in our analysis. 
\\

{\bf Hydrogen evolution reaction (HER) performance.}~In addition to the band gap and band alignments, we also evaluated 
the energetics of the hydrogen evolution reaction (HER) for the screened photocatalytic materials. To do this, we calculated the adsorption 
free energy of one hydrogen atom ($\Delta G_{\rm H}$) on the materials surface, which was obtained through the equation:
\begin{equation}
\Delta G_{\rm H} = \Delta E_{\rm H} + \Delta E_{\rm ZPE} - T \Delta S~,
\label{eq:deltaG}
\end{equation}
where $\Delta E_{\rm H}$ is the hydrogen atom adsorption energy and $\Delta E_{\rm ZPE}$ and $\Delta S$ the zero-point energy 
and entropy differences between the hydrogen atom adsorbed on the materials surface and in gas phase, respectively. The contribution 
from the catalysts to both $\Delta E_{\rm ZPE}$ and $\Delta S$ are very small and thus are neglected. The $\Delta E_{\rm ZPE}$
term was calculated like the difference between the zero point energy of the adsorbed hydrogen atom and one half of the zero 
point energy of a H$_{2}$ gas molecule, namely:
\begin{equation}
\Delta E_{\rm ZPE} = E^{\rm H}_{\rm ZPE} - \frac{1}{2}E^{\rm H_{2}}_{\rm ZPE}~.
\end{equation}
The size of the $\Delta E_{\rm ZPE}$ term is usually small and of the order of $\sim 0.01$~eV \cite{norskov05}. 
Meanwhile, $\Delta S$ was taken equal to minus one half of the experimental entropy of a H$_{2}$ gas molecule under standard 
thermodynamic conditions (i.e., $298$~K and $1$~atm), which leads to $T \Delta S = -0.2$~eV \cite{atkins14}. 
\\

\section*{Data availability}
The data that support the findings of this study are available from authors Z.L. and C.C. upon reasonable request.

\section*{Acknowledgements}
C.C. acknowledges support from the Spanish Ministry of Science, Innovation and Universities
under the ``Ram\'on y Cajal'' fellowship RYC2018-024947-I.
Z.L and B.W. acknowledge support from the National Natural Science Foundation of China (12002402, 
11832019), the NSFC original exploration project (12150001), the Project of Nuclear Power Technology 
Innovation Center of Science Technology and Industry for National Defense (HDLCXZX-2021-HD-035), and 
the Guangdong International Science and Technology Cooperation Program (2020A0505020005).
D.C. acknowledges support from the Australian Research Council Projects (DP210100879, LP190100829).

\section*{Author contributions}
Z.L. and C.C. conceived the study and planned the research. C.C. proposed the electrostatic modelling of piezo-photocatalyst 
materials. Z.L. performed the high-throughput computational screening of piezo-photocatalysts based on the Materials Project 
database. Z.L. and C.C. performed the supplementary and validation first-principles calculations and analysed the results along
with the rest of co-authors. The manuscript was written by C.C. with substantial input from the rest of co-authors.

\section*{Additional information}
Supplementary information is available in the online version of the paper.

\section*{Competing financial interests}
The authors declare no competing financial interests.

\end{document}